# Control Theoretic Approach to Predict Shock Response in Yeast

*Meenakshi Chatterjee*

Department of Electrical Engineering
Yale University, New Haven, CT 06511, USA
Email: meenakshi.chatterjee@yale.edu

*Abstract—* **This report formulates a minimal model based on a control theoretic framework to best describe the dynamics of perfect adaptation shown by the hyper osmotic shock response system in yeast. Using principles from adaptive control and stability theory, we step by step apply system identification methods to build a simple second order linear system with only a few parameters, that can concisely model the High Osmolarity Glycerol (HOG) Mitogen Activated Protein Kinase (MAPK) signaling dynamics. Validation with experimental data demonstrate that the model is sufficient to predict response of yeast to an arbitrary external osmotic shock stimulus.**

## I. INTRODUCTION

When a living cell is exposed to a medium of high osmolarity, the osmolyte concentration outside the cell becomes much higher than what it is in the inside of the cell. In such a situation, there will be a tendency of the cell to transfer water from inside the cell to the medium outside, thus resulting in shrinkage of cell volume. The osmoregulatory system of the organism is designed to help the cell adapt to the external pressure, hence balance out the pressure differences and thus prevent cell shrinkage. In Budding Yeast, the main module of the hyper osmotic shock system consists of the High-Osmolarity Glycerol (HOG) Mitogen Activated Protein Kinase (MAPK) cascade. After a hyperosmotic shock, membrane proteins of the yeast cells trigger the signal transduction cascade that results in activation of HOG1 protein. This HOG1 protein, which was initially in the cytoplasm of the cell when there was no shock, now becomes activated (due to osmotic stock) and travels to the nucleus (Fig.1). Active HoG1 in nucleus then activates a broad transcriptional response to osmotic stress (Fig.2). This means that with the inducement of osmotic stress, cells increase or decrease their export rate of glycerol (which is the osmolyte of the cell) through its transmembrane. In addition, under high osmotic stress and over longer time scales, active nuclear HOG1 also modifies the expression of some regulatory proteins that in turn regulate the production of gylcerol (thus more production of glycerol will help the cell to further effectively balance the osmolyte concentration inside and outside the cell). When osmotic balance is regained, MAPK HOG1 cascade activity ceases and HOG1 returns back to the cytoplasm. To estimate the amount of active HOG1 in living cells, the cellular localization of Hog1 protein can be monitored.

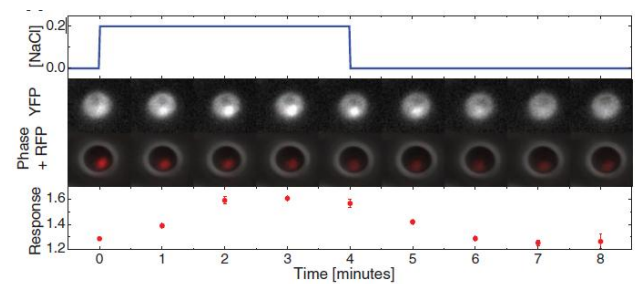

Fig1. Localization of the Hog1 protein in nucleus of cell by fluorescence microscopy. NaCl (0.2 M) was applied and removed as shown by the blue line. Figure taken from [1].

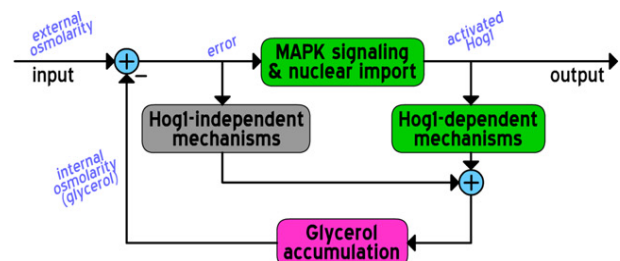

Fig2. Network diagram showing the system input and measurable outputs. Error is the deviation from optimal turgor pressure. Figure taken from [2].

## II. MOTIVATION AND OBJECTIVE

Cells are able to sense and respond to environmental changes through cascades of biochemical reactions that occur with rates spanning a wide dynamic range. A system may be composed of many parameters but only a few may be critical in inferring the signalling dynamics. Incorporating knowledge of all reactions in the system could potentially lead to a bulky model with nearly hundreds of parameters, many of which could be not be easily measurable or biologically identifiable. In this report, we consider the control problem of keeping the turgor pressure constant, given disturbance in the external osmotic pressure. We focus particularly on the MAPK HOG1 cascade as both the input (extracellular osmolyte concentration) and the output of the network (activity of MAPK HOG1 protein) are easily measured and manipulated. The osmoregulatory system of yeast contains multiple negative feedback loops that are both dependent or independent of HOG1 MAPK cascade, and operate over different time scales. An interesting aspect to address is to understand which negative feedback loop largely dominates the signal dynamics and whether the different feedback loops have distinct biological functions. Here we apply system identification methods to infer a concise predictive model of the signalling dynamics. A minimalistic model like this can excel at providing intuitive and general insights into the dynamic properties of the system.

## III. EXPERIMENTAL DATA

All experimental data has been obtained from authors in [1]. The dataset is briefly described here: cells were periodically shocked with square wave pulses of 0.2 M NaCl (Fig. 3). The response of the system to square wave stimuli with periods ranging from T = 2 minutes to T = 64 minutes was measured. In case of stimulus with T = 2 minutes, the response was measured over 10 time periods, for both T = 4 and 8 minutes it was 8 time periods, for T= 16 minutes, it was 6 periods, and finally for both T = 32 and 64 minutes, it was 4 time periods.

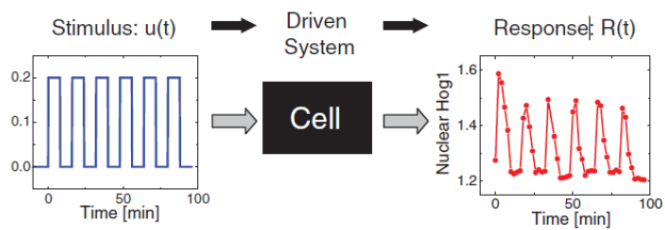

Fig 3: Response plot for a cell driven by a square wave stimulus of NaCl. Figure taken from [1].

## IV. SYSTEM MODELLING

The order of the system is not known to us. The key step is to first determine the order of the system, and then find out the value of the corresponding coefficients. As a first step to model the behaviour of the system, we start with assuming simpler functions and equations, such as for example by assuming a linear first order system and then move on to model with a second order system.

### A. FIRST ORDER SYSTEM MODEL

Let us first assume that the real system has the following dynamics:

$$\dot{x} + ax = bu \qquad (1)$$

We want to build an estimator (given below) that will accurately track the state of real system. The estimated model is:

$$\dot{\hat{x}} + \hat{a}\hat{x} = \hat{b}u \qquad (2)$$

To do so, we want the error between the estimated and actual state to converge, or in other words we can design the estimator in such a way, that the system of equation involving the error is stable. If the actual model in (1) truly describes the data, then the error between the estimated and the actual system should finally converge. We can then get the estimated values of the coefficients from the resulting model. The simulation has been performed in the following way:

$$\hat{x} - x = e \ ; \hat{a} - a = \tilde{a} \ ; \hat{b} - b = \tilde{b} \ ; \hat{c} - c = \tilde{c} \ ; \dot{\tilde{a}} = \dot{\hat{a}} = e\hat{x}$$
$$\dot{\tilde{b}} = \dot{\hat{b}} = -eu \ ;$$

The error plots corresponding to square wave input of T= 2, 4 and 8 is shown below:

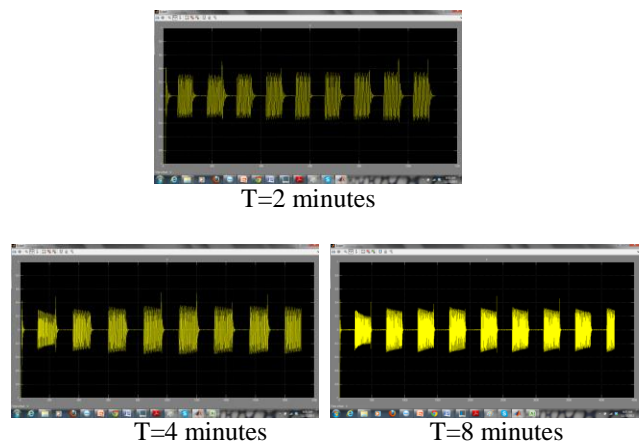

T=2 minutes

T=4 minutes      T=8 minutes

Since the errors do not converge, we can say that the real system does not follow the dynamics as given in (1).

As the second step, a simulation was performed to incorporate the effect of the derivative of the control input. The model assumed is

$$\dot{x} + ax = bu + c\dot{u}$$

The system equation of the estimated model is

$$\dot{\hat{x}} + \hat{a}\hat{x} = \hat{b}u + \hat{c}\dot{u}$$

The implementation of the simulation is as follows:

$$\dot{x} = -ax + bu + \hat{c}\dot{u} \ ; \dot{\hat{x}} + \hat{a}\hat{x} = \hat{b}u + \hat{c}\dot{u}; \hat{x} - x = e;$$
$$\hat{a} - a = \tilde{a}; \ \hat{b} - b = \tilde{b}; \ \hat{c} - c = \tilde{c}; \ ; \dot{\tilde{a}} = \dot{\hat{a}} = e\hat{x};$$
$$\dot{\tilde{b}} = \dot{\hat{b}} = -eu \ ; \ \dot{\tilde{c}} = \dot{\hat{c}} = -e\dot{u} \ ;$$

The error plots corresponding to square wave input of T= 2, 4 and 8 is shown below.

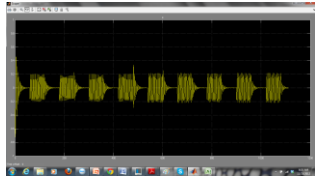
T=2 minutes

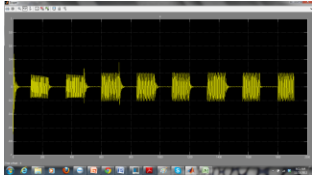
T= 4 minutes

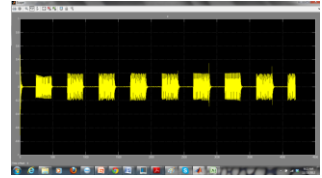
T=8 minutes

Again, we see that the error does not converge at all. Hence, we can conclusively say that a linear first order model is not rich enough to describe the dynamics of the real system. In a complicated system like this, there can be many molecules that can correspond to the states of the system. Candidate molecules that may correspond to the relevant dynamic variables should include factors known to be important in the system (e.g., glycerol and cell volume for the yeast osmotic shock response) and factors whose dynamics are expected to change on an appropriate timescale (e.g., fast protein-protein interactions). Therefore, it is not surprising that the real system is not of first order.

### B. SECOND ORDER SYSTEM MODEL

The next approach is to check if the system is of second order or not. The model assumed is:

$$\ddot{x} + a\dot{x} + bx = cu$$

The estimated model is:

$$\ddot{\hat{x}} + \hat{a}\dot{\hat{x}} + \hat{b}\hat{x} = \hat{c}u$$

Simulation of the above has been performed in the following way:

$$\dot{x}_1 = x_2;\ \dot{x}_2 = -ax_2 - bx_1 + cu;\ \dot{\widehat{x_1}} = \widehat{x_2};$$

$$\dot{\widehat{x_2}} = -\hat{a}\widehat{x_2} - \hat{b}\widehat{x_1} + \hat{c}u;\ \widehat{x_2} - x_2 = e_2;\ \widehat{x_1} - x_1 = e_1;$$

$$\hat{a} - a = \tilde{a};\ \hat{b} - b = \tilde{b};\ \hat{c} - c = \tilde{c};\ \dot{\tilde{a}} = \dot{\hat{a}} = e_2\widehat{x_2};$$

$$\dot{\tilde{b}} = \dot{\hat{b}} = e_2\widehat{x_1};\ \dot{\tilde{c}} = \dot{\hat{c}} = e_2 u;$$

The error plots corresponding to square wave input of T= 2, 4 and 8 is shown below.

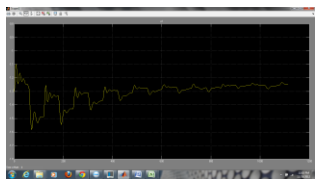
T=2 minutes

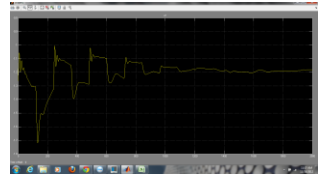
T=4 minutes

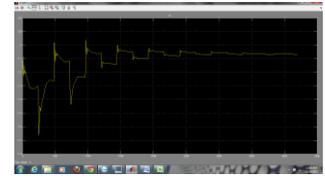
T=8 minutes

We see that although not exactly, but the error converges. The latter however converges to a value around -1.2 in all the plots, which implies that there must exist some bias in the real system. The values of coefficients obtained is as follows:

$$\hat{a} = 0.155;\ \hat{b} = 0.075;\ \hat{c} = 0.00797$$

Experimentally it has been observed that the amount of HOG1 protein in the nucleus of the cell does not significantly decrease below a basal level $R_0$, found in absence of the stimulus. From literature we know, that the value of $R_0$ for this cell type is 1.237. This might be responsible for the bias seen in the simulated system. We can easily augment the model and increase its accuracy in this scenario by passing the output of the system, through a static nonlinear element. Another second order model that was simulated is:

$$\ddot{x} + a\dot{x} + bx = c\dot{u}$$

The estimated model is:

$$\ddot{\hat{x}} + \hat{a}\dot{\hat{x}} + \hat{b}\hat{x} = \hat{c}\dot{u}$$

The simulation was performed similar to the previous case. The error plots corresponding to square wave input of T= 2, 4 and 8 is shown below:

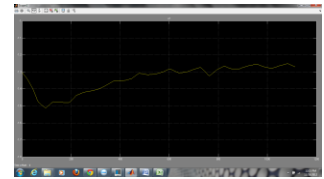
T=2 minutes

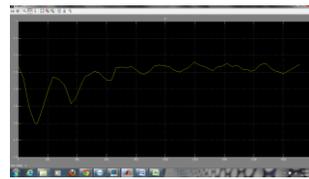
T=4 minutes

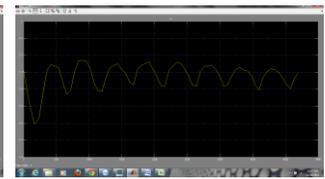
T=8 minutes

We again see that the error converges, although not accurately. The latter however converges to a value around -1.2 in all the plots. The values of coefficients obtained is as follows:

$$\hat{a} = 0.14;\ \hat{b} = 0.001;\ \hat{c} = 0.1002;$$

## C. MODIFICATIONS OF THE SECOND ORDER SYSTEM MODEL

To improve or fine tune the error convergence, some modifications are made on simulation of the second order system. The model assumed is:
$$\ddot{x} + a\dot{x} + bx = cu$$

The estimated model is:
$$\ddot{\hat{x}} + 0.2\dot{\hat{x}} + 0.1\hat{x} + (\hat{a} - 0.2)\dot{x} + (\hat{b} - 0.1)x = \hat{c}u$$

The implementation is as follows:
$$\ddot{\hat{x}} + 0.2\dot{\hat{x}} + 0.1\hat{x} + (\hat{a} - 0.2)\dot{x} + (\hat{b} - 0.1)x = \hat{c}u;$$

$$\dot{x}_1 = x_2;\ \dot{x}_2 = -ax_2 - bx_1 + cu;\ \dot{\widehat{x_1}} = \widehat{x_2};$$

$$\dot{\widehat{x_2}} = -0.2\widehat{x_2} - (\hat{a} - 0.2)x_2 - 0.1\widehat{x_1} - (\hat{b} - 0.1)x_1 + \hat{c}u;$$

$$e = \hat{x} - x;\ \hat{a} - a = \tilde{a};\ \hat{b} - b = \tilde{b};\ \hat{c} - c = \tilde{c};$$

$$\ddot{e} + 0.2\dot{e} + 0.1e = -\tilde{a}\dot{x} - \tilde{b}x + \tilde{c}u;\ \epsilon = 0.5e_1 + 9e_2;$$

$$\dot{\tilde{a}} = \dot{\hat{a}} = \epsilon x_2;\ \dot{\tilde{b}} = \dot{\hat{b}} = \epsilon x_1;\ \dot{\tilde{c}} = \dot{\hat{c}} = \epsilon u;$$

The error plots corresponding to square wave input of T= 2 is shown below.

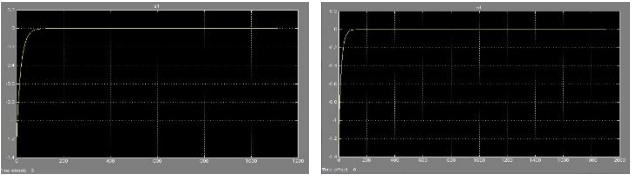

T= 2 minutes          T=4 minutes

The values of coefficients obtained is as follows:
$$\hat{a} = 0.1995;\ \hat{b} = 0.0825;\ \hat{c} = 0.1025$$

## D. DIRECTIONS FOR MORE COMPLEX MODELS

If the system were composed only of reactions that could be modelled with linear dynamics, then such a result might be close to trivial. However, the fact that the experimentally obtained peak Hog1 amplitude saturates as a function of salt (Fig 4) is strong evidence of nonlinearity in the system.

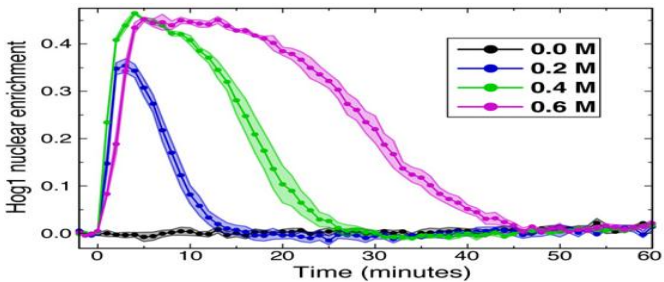

Fig4. Response of cells treated with hyper-osmotic shocks with indicated concentration of NaCl. Figure taken from [2]

Thus, an alternative model would be to incorporate nonlinearity in a first order system and it will be interesting to observe how the system responds. The system model could be like:
$$\dot{x} = \beta_1 x + \beta_2 x^2 + \beta_3 x^3 + c_1 u$$

Such a system is nonlinear in its dynamics and linear in control. Additional extensions on this model can lead to models like:
$$\dot{x} = \beta_1 x + \beta_2 x^2 + \beta_3 x^3 + c_1 u + c_2 u^2 + c_3 u^3$$

## V. DISCUSSION AND CONCLUSION

Minimalistic modelling approach in Systems Biology literature mainly involve following the traditional way of parameter constraining, performing observability tests and then estimating the parameters, or doing analysis in the frequency domain. Such models can then be used to predict response to a step input and then compared with the experimental data to test its validity. Similar models can be implemented for those cells which has been genetically engineered to have a very weak MAPK cascade activity and their response can be observed to a step input. Comparison of the amplitude of this response and the time taken to reach the basal level to that of the cells whose MAPK cascade is active, can shed light on the role of HOG1 dependent mechanisms in osmoregulation. Here we see that a basic result from control engineering can successfully provide intuitive information about biological mechanism. Similar analyses can be useful in other homeostatic systems (e.g., blood calcium levels), showing perfect adaptation. Identification of biological mechanisms responsible for driving perfect adaptation will be critical in studying homeostatic systems and the designing of perfectly adapting synthetic circuits in future.